Published in {\it Experimental Metaphysics-Quantum Mechanical Studies in Honor of 
Abner Shimony}, volume 1, edited by R. S. Cohen, M. A. Horne and J. S. Stachel 
(Kluwer, Great Britain 1997) pp.143-156. 

\bigskip
\font\big=cmr17
\font\title=cmbx12
\vglue 1 cm
\centerline{\big{Tales and Tails and Stuff and Nonsense}}

\vskip 1cm
\centerline {Philip Pearle}
\centerline{\it{Department of Physics, Hamilton College, Clinton, NY  13323}}
\bigskip
\centerline{\tenbf{Abstract}}

{\tenrm In an informal way I review collapse models and my part in constructing them, 
and I recall some encounters with Abner Shimony.  In particular, I address the question of the 
nature of spacetime reality in collapse models, stimulated by Abner's
criticism of the ``tail" possessed by statevectors in such models.}
 
\vskip 1cm
\leftline {\title 0. Prehistory}
\bigskip

	Normally, when writing Physics Prose, one is expected to expunge all
indications that physics is a human enterprise.  But that 
expectation may cheerfully be waived in a volume honoring Abner Shimony.  
No one else embodies for me such a graceful blend of physics and kindly humanity.
So I will take advantage of 
this opportunity to reminisce on the physics enterprise which 
has been the major work of my life, and weave in some tales 
of some of my interactions with Abner over the years 
which have been important and memorable to me.  I also want to particularly
focus on an issue called ``tails." Ten years ago,
Abner took, and continues to take on this issue, a position with which 
I once agreed, but now I am in apostasy.  In so doing, I will discuss
the larger issue which is raised, namely
the nature of
the spacetime reality given by collapse models.

	I first got to know Abner 29 years ago.  We were introduced by Wendell Furry,
in his office in the Jefferson lab at Harvard where I had my first job after
getting my Phd. degree. I was working
on my first paper,$^{1}$ which 
was critical of Standard Quantum Theory (SQT). From my undergraduate days
I simply could not believe that SQT was a complete picture of nature.
I couldn't understand
why my teachers and the rest of the physics community conveyed so 
forcibly the order ``don't question it, just use it".

  After a good deal of
agony I decided not to obey that order. 
In that paper, among other things, I concluded that the Collapse Rule, the abrupt 
replacement of a statevector (hitherto smoothly evolving
via Schr\" odinger's equation) by a projection of that statevector was ill
defined because no one could tell you precisely the circumstances 
in which it should be applied, nor precisely the time of application.
Wendell was the only person I had found who was sympathetic to 
these concerns.  In Abner I found
another.  I remember hearing for the first time about the Everett interpretation
from him, 
which he pointed out had some features in common with some aspects of my
discussion.  But mostly I remember my delight at receiving his warm
encouragement, and the realization that it was possible to earn a living 
working at Foundations of Quantum theory.  After all, I had a data point---Abner.

	Well, maybe it was possible to earn a living 
working at Foundations, but it wasn't possible for me to do it at Harvard.  
A few years later, Abner came to visit Howard Stein at Case Institute of Technology,
where I had my second job.  I vividly recall a conversation we had in the open
courtyard.  Abner said he was interested in seeing if an experiment
with photons could be devised which would test Bell's inequality, and asked if 
I was interested in working upon it.  I gave that proposal the deep consideration
that it required---for one second---and answered {\it no}!  My reason, I explained, 
was that I knew how the experiment would turn out, in favor of quantum theory
and in opposition to local hidden variables.$^{2}$ 
For some reason, in spite of my
argument, Abner did not abandon his idea  and the result was the celebrated paper by Clauser,
Holt, Horne and Shimony.$^{3}$  What did I conclude from this?
\vskip 12 pt
 	I was right!  I was wrong!
\vskip 12 pt	
 	In normal fields of human endeavor it would be impossible to 
reconcile such a conflict, but fortunately we are in Foundations and so we
have recourse to
\vskip 12 pt
\centerline {\bf Niels Bohr's Unabridged Dictionary}
\centerline {(first edition, Random House)}

\hskip 1 in {\bf truth}\qquad (tr$\overline {oo}$th), n.; pl. TRUTHS
(tr$\overline {oo}$ths) 

\hskip 1 in [Middle English trewthe, trowthe: from Anglo Saxon treowthu]

\hskip 1 in 1. ordinary truth: opposite is false

\hskip 1 in 2. deep truth: opposite is also deep truth
\vskip 12 pt

\noindent Clearly Abner had given me some personal insight into deep truth.
I concluded

$${\rm You\  should\  always\  listen\  to\  Abner!} \eqno {(0.1)}$$

	Actually there is no need for deep truth (even in his humorous moments, 
Niels Bohr seems to have had a predilection
to do public relations work for illogic, i.e., non-sense). I was right about
one thing but wrong about another.  I was right about the experimental result.
But I was wrong not to get involved.  I learned that one can {\it think}
one knows the result of an experiment, but it is only after the experiment
has been performed that one {\it knows} one knows the result.  I also learned,
by watching Abner at various conferences, how much {\it fun} it is, 
especially for a theorist, to go around talking about experiments.

	Well, maybe it was possible to earn a living 
working at Foundations, but when Case turned
into Case Western Reserve it wasn't possible for me to do it
there, and in 1969 I went to Hamilton 
College, where it is possible (it is worthy of remark that an unusually
large number of people in this field have found a home at
small liberal arts colleges).

	 At Hamilton I wrote a paper$^{4}$ on a way of
getting around Bell's inequality if the particles in an EPR-Bohm experiment
have three options: not only may they respond as spin-up or spin-down to 
the apparatus, but they can also elect not to be detected.  If the experimenter 
throws away the data where the particles don't appear, I showed by a local
hiden variable model 
that one can then get from the remaining data 
at each angle the correct quantum correlations.  This
possibility has come to be called ``the loophole."
I gave a talk about this at a conference in London, Ontario in 1971,
and Abner gently informed
me from the audience that an undesireable feature in my model
would surely be spotted by an experimenter. It is that the number
of detected particles is not constant, but depends upon the angle between
the spin detectors. (This defect was later corrected in a model of Clauser
and Horne.$^{5}$) I had also worked out a model-independent bound of 14\%
on the fraction of particles which would go undetected (the best
bound I know of now, due to Mermin,$^{6}$ is 17\%),
and concluded in my paper ``\dots
it is difficult to take this hypothesis seriously as a physical principle
capable of extension to a large group of phenomena: had such large fractions
of undetected events occurred in other already performed correlation 
experiments, it is hard to see how such behavior would have gone unnoticed."
I recall how flabbergasted I was to learn from Abner
that photon counters are so inefficient that such large fractions of undetected
events occur all the time.  Only now, two decades later, is an 
experiment in the works that seems likely to close the loophole (see Ed Fry's 
paper in this volume).  I {\it think}
I know what the result of this experiment will be, but \dots.

\vskip 1cm
\leftline {\title 1. Early collapse models and Gambler's Ruin}
\bigskip

	Just about this time I had an idea of how to go about correcting what 
I perceived to be the Achilles heel of SQT,
the Collapse Rule. In 1966 Bohm and Bub published a paper$^{7}$
in which they describe the collapse of the statevector as due
to its interaction with Wiener-Siegal hidden variables.  The hidden variables 
(which any system is supposed to possess with a given probability distribution)  
guide the superposition of states to evolve (collapse) to one of them.  
The probabilities and evolution are such that the ensemble of all collapsed states
is identical to the ensemble obtained by applying the Collapse Rule. This
showed that a modification of Schr\" odinger's equation to describe
collapse as a dynamical process, instead of an ill-defined
postulate, was possible.

  However, I felt that the Wiener-Siegal hidden variables 
had more mathematical than physical appeal, especially because 
you have to, in effect, collapse 
{\it them} to new values every time a statevector collapse is completed.
Bohm and Bub made a suggestion about the characteristic time for 
this hidden variable evolution, but
they never actually wrote down an equation that would do the job.  It 
seemed an unnatural and unnecessary complication. 
	
	I thought that it is more natural for some
randomly fluctuating quantity to be responsible for driving the statevector 
in its collapse.  After all, many things in nature fluctuate---presumably
one could eventually identify the appropriate candidate. But, best of all,
I had a simple way for the fluctuating quantity to do the job.  I will first describe
the mechanism here by a precise analogy I stumbled
upon only a decade later.$^{8}$ The analog is the Gambler's Ruin game.

	Consider two gamblers, $L$ who starts with, say, \$30 and $R$ who starts with
\$70. This corresponds to the statevector

$$|\psi,0>=\sqrt {.3}|L>+\sqrt {.7}|R> \eqno {(1.1)}$$

\noindent They toss a coin (= random fluctuation!).  Heads, $L$ wins \$1 
from $R$: tails, $R$ wins \$1 from $L$.  The money each possesses fluctuates, 
although the sum is of course constant. This corresponds to the 
evolving statevector 
	
$$|\psi,t>=c_{L}(t)|L>+c_{R}(t)|R> \eqno {(1.2)}$$

\noindent where the coefficients $c_{L}(t)$ and $c_{R}(t)$ likewise fluctuate, however
keeping the sum of their squares constant.  Eventually, one or the other
gambler wins all the money, and the game stops.  This corresponds to 

$$|\psi,t>\lower8pt\hbox{\overrightarrow{t\to \infty}}|L>\hskip 1cm {\rm OR}\hskip 1cm
|\psi,t>\lower8pt\hbox{\overrightarrow{t\to \infty}}|R> \eqno {(1.3L,R)}$$

\noindent The punch line is that, if the game is repeated many times, $L$ wins 30\% of 
the games and $R$ wins 70\%. This corresponds to (1.3L) occurring .3 of the
time, and (1.3R) occurring .7 of the time, just the probabilities 
assigned by SQT's Collapse Rule to the $|L>$ and $|R>$ outcomes respectively!

	The proof of this is really simple.  Let $P(x)$ be the probability
that, say, gambler L wins the game when he possesses the fraction x
of the total money.  Then after the next toss there are two possibilities:
either he loses the toss and then wins the game or he wins the toss and wins the 
game, so

$$P(x)={1\over2}P(x-.01)+{1\over2}P(x+.01)\eqno {(1.4)}$$

\noindent The solution of this difference equation, satisfying the obvious 
boundary conditions $P(0)=0$ and $P(1)=1$, is $P(x)=x$.

	A variant of this game is that when one of the gamblers is down to his last dollar 
they agree to play for 50 cents, and if he loses that they play for 25 cents, etc., so
the game never ends.  In the collapse dynamics, this means that a coefficient
in Eq.(1.2), say $c_L(t)$,
can get very small (and, it should be emphasized, with probability $1-|c_L(t)|^2$ it will continue to 
get smaller yet), but it never vanishes. Such a small piece of the statevector
is called a `tail', and I will be saying a lot more about it later on.

	The strategy I seized upon was to look for a modification of Schr\" odinger's
equation so that the ensemble of solutions obeys a
diffusion equation which embodies (what I later learned is) Gambler's Ruin 
behavior.  For example, for the two-state system (1.2), set $|c_L(t)|^2=x$ and suppose the
ensemble of x-values obeys the diffusion equation

$${\partial \rho(x,t)\over \partial t}=\lambda {\partial^{2}\over \partial x^{2}}
x(1-x)\rho(x,t)\eqno {(1.5)}$$

\noindent ($\lambda^{-1}$ characterizes the collapse time) with initial condition
 $\rho (x,0)=\delta (x-x_{0})$ .

	It is easy to see from Eq.(1.5), without solving it,
that it provides the essential aspects of the Gambler's Ruin game. 
From integration of Eq.(1.5) 
over $x$ from from $0-$ to $1+$ (where $\rho$ vanishes) one 
learns that $\bar \rho\equiv \int \rho dx$ is constant (at value 1) in time.
Likewise, multiplication of Eq. (1.5) by $x$ followed by
integration over $x$ leads (after an integration
by parts) to the 
result that $\bar x\equiv \int x\rho dx=x_{0}$.  This constancy of $\bar x$
implies that the ``game" is fair, 
or what mathematicians
call a Martingale.  Similarly, multiplying by $x(1-x)$, one finds that
d$\overline {x(1-x)}$/dt$=-2\lambda \overline {x(1-x)}$, so 
$\overline {x(1-x)}=x_{0}(1-x_{0})\exp -2\lambda t$.

Since $\rho$ and $x(1-x)$
are positive or zero, the only way that $\overline {x(1-x)}$ can vanish as $t\rightarrow\infty$
is for $\rho$ to vanish where $x(1-x)$ doesn't, and vice versa.  This means that asymptotically
there is collapse: $\rho\rightarrow  a\delta (1-x)+(1-a)\delta (x)$ 
($a$ is some constant: the sum of coefficients of the delta functions must add up to
1 since $\overline {\rho}=1$). But, because of the Martingale property, $\bar x(\infty)=a$ must in fact
be the initial value $x_{0}$. Therefore $\rho\rightarrow  x_{0}\delta (1-x)+(1-x_{0})\delta (x)$
and we have (a continuous-in-time version of) the complete
Gambler's Ruin behavior.  It can be shown$^{9-11}$ that the tails behavior, 
a game which never ends, is obtained if $x(1-x)$ in Eq. (1.5)
is replaced by $[x(1-x)]^{r}$, where $r\geq 2$.

	The only problem was to find
a modified Schr\" odinger equation whose ensemble of solutions would
obey the diffusion equation. 
I went on my first sabbatical in 1973 to the University of Geneva, and
both Josef Jauch and John Bell (with whom I had corresponded
but had not previously met) were encouraging.  But it was only a year later
that I found tools in a book of Prigogine$^{12}$ and a paper of 
Chandrasekhar$^{13}$ which enabled me to
try out various modified Schr\" odinger equations and find the diffusion equations
they imply. 

	Some weeks after submitting my first paper$^{14}$ on this subject,
 I received a telephone call from Fred Belinfante.  He told me that he had been sent
the paper to referee, and that he liked it very much.  But he said that there was a prejudice
against such work in the offices of Physical Review D, and he didn't want to criticize
the paper in any way in his report, for fear it would be used as an excuse not to publish.
He had some questions and some criticisms, and so we talked about them.
Not long thereafter I was informed that the paper had been
accepted, and I made a few revisions to respond 
to his concerns.  For a while that prejudice abated, but in the last six years 
it has come back.  A
special set of rules has been framed, aimed at `` 
speculative theoretical papers that lie outside the mainstream of
current research," which emphasizes that such
work, to be considered for publication, must have experimental consequences.  The fact that
this work {\it does} have such consequences, and 
a great deal of what they publish does not, doesn't seem to affect their obduracy.

	The equation I proposed was nonlinear in $|\psi ,t>$: 
	
$$	i{d\over dt}<a_{n}|\psi, t>=<a_{n}|H|\psi, t>+\lambda
\sum_{m}<a_{n}|A|a_{m}>{\biggl[<\psi,t|a_{m}><a_{n}|\psi,t>\biggr]^{r}\over<\psi,t|a_{n}>}\eqno{(1.6)}$$

\noindent  I give it
 here, although a far superior formulation will be presented later,
because it illustrates all the issues which had and have to be struggled with.

\vskip 1cm
\leftline {\title 2. Issues}
\bigskip

	In this first paper, I assumed H and the initial 
conditions are such that the phases of $<a_{m}|\psi,t>$
are essentially random.
My argument was not so clean because I had not yet
encountered the powerful methods of Ito for handling nonlinear
differential equations which are linear in white noise. Then I came across a book by Wong$^{15}$
which taught me just what was needed. It was possible to write a paper 
with a much neater argument,$^{16}$
with the matrix elements $<a_{n}|A|a_{m}>$ in Eq. (1.6) set
proportional to (complex) white noise (for those who care, Eq.(1.6)
is then a Stratonovich stochastic d.e.). 

	White noise, given ab initio, has been
used to drive collapse equations
ever since.  But I have always thought of this as a mathematical convenience: nothing in 
the physical world has the same behavior at all frequencies.  Indeed, only recently 
I have been driven, by the need to get rid of infinities
in relativistic collapse models, to consider non-white noise mechanisms$^{17, 18}$ 
and I
regard being forced by the physics to adopt a possibly more
realistic noise source as a hopeful sign.
However, ultimately I think that the noise will turn out
to be an approximation to some more dynamical and interactive quantity.

  As to the issue of
making a connection between the collapse mechanism and the rest of physics---what is the
 physical identity of the white noise?---gravity
has presented itself as the leading candidate.$^{19-23}$ At the time I thought
gravity was likely (in ref. 14 I gave an example with $\lambda^{-1} = 10^{19}$sec
involving an apparatus
whose collapse time was $10^{-15}$ sec, and
remarked that $\hbar \lambda \approx 10^{-8}$erg, 
comparable to the apparatus's gravitational self-energy: I concluded ref. 16
with a homily
on the involvement of gravity), on the not very profound
principle that what is understood least has the greatest
potential for surprises and might therefore be most responsible for
collapse. I spent my second
sabbatical in 1981 at Oxford with Roger Penrose's group, for
very early Roger had argued for a connection between gravity and collapse and he has been
active in suggesting some tentative physical principles.$^{20}$ 
But think it is fair to say that the formulation of this connection is only in a rudimentary stage.

	A word now about simplicity.  Unfortunately, we have not yet even {\it one}
experiment whose outcomes disagree with the predictions of SQT to guide us in 
constructing collapse models---except that events occur in {\it every}
quantum experiment and SQT's crude handling of them indicates the need for a revised quantum theory.
The constraints on
a viable collapse model are enormous: to agree with {\it every} experiment
while describing collapse.  It always has been my hope that these constraints
and Dirac's dictum to look for simplicity in one's equations would provide
sufficient guidance to this program for at least a while, and that attractive
physical principles would eventually emerge along with constraining experiments, and I believe
it has worked that way so far.  However, simplicity is in the eye of the
beholder.  

	Which brings me to tails.  I had favored Eq.(2.6) with $r=1$ largely 
on the grounds of simplicity, but I wanted a physical principle which
would back up this choice.  In 1984 Abner invited me to speak at Boston University,
and stay overnight at his house.  We spoke in his study about choosing $r=1$ with
no tails or $r=2$ with tails, and Abner convinced me (recall principle (0.1)!) that, as I
put it in a paper (written because of this interaction with Abner) 
on the collapse time,$^{9}$ ``\dots the reduction
must take place in a finite time because an experimental result reveals itself in a finite time.''
I took this position in subsequent papers$^{10, 24}$ too, that the presence
of a piece of the statevector that corresponds to another outcome of
an experiment, no matter how small compared to the piece that describes the actual
outcome, is ontologically undesireable. 
In this paper I will eat my words.

	There were two other problems which I could not figure out how to handle. From the
first I mentioned them,
in every paper and every review$^{11,24-26}$: I simply didn't know what to do.
	
	One I call the Preferred Basis problem. In Eq.(2.6) the basis states $|a_{n}>$ to which 
collapse takes place are put in by hand, and are not really described in any precise way.
This is identical to, and therefore as
bad as, SQT's inability to say which states end in collapse except in an ad hoc way.
To be sure, I had some ideas,
suggesting$^{13}$ that it ``\dots depends only upon the the macroscopic distinguishability
of appropriate position variables characterizing the states.''  But the preferred 
states couldn't be position eigenstates because that would require the collapse process
to impart infinite
energy to them. 

	The other I call the Trigger problem. The operator $A$ in Eq.(1.6) should be small
for states that differ only in some microscopic way, since one doesn't want e.g., two separated
packets belonging to a single particle to collapse to one packet in a short time, since
that would mean the theory predicts no interference will take place when Nature
says it does.  On the other hand, one wants rapid collapse when the two packets describe
a macroscopic apparatus variable in two different places.  How do you make the additional
term small for microsystems, but large for macrosystems?  I just did it by hand, but 
that was hardly a satisfying solution to the problem.

\vskip 1cm
\leftline {\title 3. GRW's Spontaneous Localization (SL) Model}
\bigskip

	Then along came the work of Ghirardi, Rimini and Weber.$^{27}$  They didn't know
about my work,they didn't have a modified Schr\"odinger equation, but they solved
my two problems. Actually, I hadn't appreciated what they had done when I first
read their paper. I was so misled by their emphasis on the density matrix's (neat!) behavior that I 
did not grasp their handling of the collapse mechanism for the individual statevector. It wasn't
until I read a preprint by John Bell$^{28}$ that
I understood what they had accomplished. The image of
a jigsaw puzzle with only two missing pieces announcing their shape
by their holes came to my mind, with the pieces suddenly found and snapped into place,
so clever and convincing were their ideas, and so tuned were they to what I wanted but hadn't
been able to find.

	To understand GRW's model, consider first a particle described by two 
packets in one dimension.  For definiteness, 
let $\psi (x)=\phi (x)+\phi (x-l)$, where $\phi (x)$ is centered on
$x=0$ with width much less than $a\equiv 10^{-5}$cm and $l$ much greater than $a$ ($a$ is
one of two parameters in the model).  If we suppose the Hamiltonian is zero, the wavefunction
will just sit there until \dots WHAM, a HIT occurs.  That is, the wavefunction is suddenly multiplied
by a gaussian, say exp$-(x-.1a)^{2}/2a^{2}$, and immediately renormalized to 1. 
The result of this is that the left packet is slightly altered
in shape, but boosted in amplitude, and the right packet is almost wiped out. 
A collapse has occurred,
with a small tail remaining.

	I have deliberately chosen the center of the gaussian
to be near the peak of one of the packets because that is a probable thing to
happen.  Along with the dynamical behavior, the model specifies a Probability Rule.  It
is that the probability that a hit occurs in the time interval $(t, t+dt)$
and that the center of the hit lies in $(x, x+dx)$ is $\lambda N^{2}dxdt$, 
where $N$ is the norm of the wavefunction after the hit but before renormalization.
If the center of
the gaussian is far from both packets, $N$ is very small, so that is very improbable.
The parameter $\lambda ^{-1}$ was chosen by GRW to be around $10^{16}$ sec $\approx$ 300
million yrs, so a microscopic system is seldom affected.  Then, why bother?

	Here's why.  Let the wavefunction now describe an apparatus ``pointer", containing $n$
particles, in a superposition of two spatially separated states. If
{\it any} particle is hit, the {\it whole} wavefunction is collapsed to one
pointer (and a tail).  Moreover, each particle 
in the apparatus has an independent chance to be hit, which makes the mean time
for a hit to occur equal to $(\lambda n)^{-1}\approx 10^{-7}$ for a pointer containing
$10^{23}$ particles.  

	So, the GRW solution to the Trigger Problem is that the mechanism
is always on, but it is slow when the superposition differs in the
positions of just a few particles, and fast when there are many particles
with different positions.  The Preferred Basis {\it is} position, but
it is modulo $a$ so as not to squeeze the wavefunction too much and impart too much energy.

	The GRW model is {\it very} nice, but it isn't a modified Schr\"odinger equation. Also, 
the hitting process destroys the symmetry of the wavefunction.  Since
they had solved my two problems, I wondered if I could combine what I had been doing with 
what they had done.  

\vskip 1cm
\leftline {\title 4. Continuous Spontaneous Localization (CSL) Model}
\bigskip

  I was on my third sabbatical in 1988, and through the intermediacy of John Bell
it was arranged that I would first spend three months in Trieste with GianCarlo Ghirardi
and, after that, a month and a half in Pavia with Alberto Rimini.  GianCarlo and I 
became friends right away. No one else, I think, believed so much in the
essential correctness of this approach as John Bell and we did.  
We spent the time trying to construct a 
relativistically invariant density matrix evolution describing collapse, with indifferent
success.  However, along the way, I tried to get the SL density matrix evolution equation
from a modified Schr\"odinger equation.

  One day, in spite of my experience
with nonlinear collapse terms, I said to myself ``if
it's fundamental, it ought to be simple, and linear is simple", and I started
looking for a way to get a linear term to describe collapse.  To my amazement, 
once I had asked the question, a scheme appeared that seemed to have promise.
However, GianCarlo and I were so focused on our relativistic problem that I put it
away for some other time.  Near the end of my stay we visited John Bell in Padua
where he was giving a lecture. Afterwards we spoke with John about our difficulties with
relativistic collapse, and also I mentioned this new idea.  However, in trying to 
explain it, I realized that I didn't understand it very well, and I determined
to work on it first chance I got. 

	That chance came when I visited Alberto, who graciously
didn't object to my working on it, showing mild polite interest at first, which turned
into strong interest as the work neared completion. As for myself, I never had an experience
with a problem which was as wonderful as this.  Everything I calculated turned to gold.
Every equation simplified, every question I asked had stimulating answers.  The whole thing took
seventeen days, from start to completed paper. I would calculate during the day, write
up the results at night, and my wife Betty would come to the University with me each day
and type it up on the computer.  Each late afternoon as we left the University
and took the long walk through the
streets of Pavia I kept repeating that I could not believe that this was happening to me. A few years
ago I heard Roger Penrose give a talk at Syracuse University, arguing that mathematics is not
invented, but it is
out there, like Platonic Ideals, waiting to be discovered.  I had felt that way,
that I was uncovering something which had been waiting.

	When I had finished the paper$^{29}$ (which subsequently languished for 4 months
at Physical Review D awaiting their
development of the ``outside the mainstream'' criterion for rejection), 
GianCarlo came to visit Alberto, as they 
both had a meeting in Milan. Apparently they  spent the  
weekend working on my paper because on Monday we met, and they had some
very nice things to say about it, and also some very interesting results
of their own which followed from it.  We decided to join some of my results
with theirs and write a paper together.$^{30}$  Alberto asked me what I wanted
to call the new model. I thought deeply---for one second---and because I was combining
my concern with continuous collapse evolution with their SL model I replied, ``Continuous
Spontaneous Localization."  Alberto said ``But," GianCarlo said ``Shh," and CSL it is.

	Without any more ado, here is the structure of CSL.  The evolution equation is
	
$${d\over dt}|\psi,t>_w=-iH|\psi,t>_w-{1\over 4\lambda}
\int d{\bf x}[w({\bf x},t)-2\lambda A({\bf x})]^{2}|\psi,t>_w \eqno (4.1)$$

\noindent I have to explain what is in (the Stratonovich) Eq.(4.1). $\lambda ^{-1}$ is GRW's
characteristic time. $w({\bf x},t)$ is any classical field: you put it in, and solve
for the statevector $|\psi,t>_{w}$ which evolves under it.  The operator $A({\bf x})$
is essentially the particle number operator for a sphere of radius $a$ (= GRW's length
parameter) centered on the point x:

$$A({\bf x})\equiv\int d{\bf z}\xi ^{\dagger}({\bf z})\xi ({\bf z}){1\over (\pi a^{2})^{3/4}}
e^{-{1\over 2a^{2}}({\bf x}-{\bf z})^{2}}\eqno (4.2)$$

\noindent ($\xi ({\bf z})$ is the annihilation operator for a particle at ({\bf z})).
It is to $A({\bf x})$'s joint eigenstates that collapse tends (any other set of commuting
operators in place of the A's would drive collapse to {\it their} joint eigenstates). Incidentally, it
is this choice of A which ensures that the symmetry of the wavefunction is
maintained during collapse.  Note that the evolution is not unitary, 
so the norm of the statevector generally changes as
it evolves.

	You can put any field $w({\bf x},t)$ you wish
into the evolution equation (4.1), but what is the probability that Nature chooses that
field? In addition to Eq.(4.1), we must specify a Probability Rule: 

$${\rm Prob}[w({\bf x},t)]=Dw\ _{w}\negthinspace\negthinspace<\psi,t|\psi,t>_{w}\eqno (4.3a)$$

\noindent  where $Dw$ is the functional integration element
   
$$Dw\equiv C\prod_{{\rm all}\ {\bf x},\ t'=t_{0}}^{t} dw({\bf x},t')\eqno (4.3b)$$

\noindent(C is chosen so that the total probability in (4.3a) is 1).
 Eq.(4.3) says that statevectors with the largest norms
are most likely to occur.

That's it!  All that is left to do is draw the consequences.  There are many ways to see
how CSL works, but we will just give one illustration.  One can think of CSL as
providing ``little hits."  Eq.(4.1), applied to a single particle in one dimension, becomes:

$${\partial \psi (x,t)\over\partial t}=\int dzw(z,t){1\over(\pi a^{2})^{1/2}}
e^{-(x-z)^{2}\over 2a^{2}}\psi (x,t)-\lambda \psi (x,t)\eqno (4.4)$$

\noindent which may be written as

$$\psi (x+dx,t)=\psi (x,t)+\int dzdB(z,t)G(x-z)\psi (x,t)-\lambda dt\psi (x,t)\eqno (4.5)$$

\noindent ($w\equiv \partial B/\partial t$). Eq. (4.5) says that, in the time interval dt, there are two terms which
modify the shape of $\psi (x,t)$.  There is added to $\psi (x,t)$ what amounts
to a hit of infinitesimal amplitude: say that in a two packet situation, 
the left packet is thus augmented by an infinitesimal amount.  There is subtracted
an infinitesimal amount proportional to $\psi (x,t)$: both packets decrease slightly.  The 
net result in this example is an infinitesimal relative increase in the size of the
left packet over the right packet. Instead of a sudden hit, 
the packets are playing the Gambler's Ruin game.

	In anticipation of the rest of this paper, it should be emphasized here that, just
as GRW's SL model leads to tails, so does CSL: this Gambler's Ruin game is of the 
never-ending variety. 

	Often when a new development in physics takes place there is a new development in
mathematics that goes along with it.  In this case, abstracting to general operators A, 
what was found was a linear statevector evolution equation describing
collapse which gives rise to the most general
Lindblad (Bloch) evolution equation for the density matrix.$^{30}$  Just before I left Pavia, Alberto
received a paper$^{31}$ from Nicolas Gisin who obtained, in the abstract case, a nonlinear 
evolution equation (essentially what we got when we combined Eqs.(4.1) and (4.3))
for the statevector which did the same thing.  
A year later, Nicolas brought to my attention a paper by Slava Belavkin$^{32}$
who had arrived at the linear statevector evolution equation in the abstract case,
 in a context totally different from 
collapse models: he was modeling continuous incomplete measurements in SQT. I wonder why we
 (see also the work by Lajos Diosi$^{33}$)
came across the same mathematical structure at almost identically the same time: 
it sometimes seems
as if ideas have their own lives, independently of people. 

\vskip 1cm
\leftline {\title 5. Spacetime Reality}
\bigskip
{\tenrm In the beginning, Schr\"odinger tried to interpret his wavefunction as giving 
somehow the density of the stuff of which the world is made.}
$$\ \eqno\hbox{\tenrm John Bell}^{34}$$
\bigskip
	At Erice in 1989, Alberto and GianCarlo reviewed SL and CSL$^{35}$
and I spoke about attempts at a
relativistic CSL.$^{36}$  John Bell gave a marvelous criticism of the flaws in SQT,$^{34}$
indicting it sarcastically
with the useful acronym FAPP (For All Practical Purposes). John was in fine fettle (at the
discussion following his talk, 
when David Mermin spoke up in defense of his friend and colleague
Kurt Gottfried's position on SQT which 
John had criticized, John snapped ``FAPP-trap"). One evening at a festa, with Abner,Alberto,
GianCarlo and others around, I raised the issue of tails to see what John would say,
and I got roundly teased for my pains. We spoke about ``stuff," the word he had recently coined
to characterize the squared wavefunction in configuration space 
$|\psi ({\bf x}_{1},\dots)|^{2}$, which he saw
as the appropriate quantity to sustain Schr\"odinger's original vision of reality, 
at last feasible because of collapse models.

 	The last time Abner, David and I saw John, at a conference
in Amherst (another small liberal arts college involved in Foundations!) in 1990,
I characterized his teasing (I had spoken of tails as ``what {\it is} contains 
a little bit of what might have been'') as a warning ``not to express a new idea in an old
language," and John smiled and nodded broadly.  His last gift to me was in his summation of 
the conference, when he said ``The GRWP theory is the most important new idea in Foundations since
I entered the field" (surely Bell's Inequality deserves mention here?---but that was typical of John's 
generosity). A few months later GianCarlo called me to tell me that John
had died of a stroke.  Like many others, we miss him greatly and we think and talk of him frequently.

	At a memorial session in 1990 in Minneapolis to honor John, GianCarlo and I discussed nonrelativistic
and relativistic CSL$^{37}$ and Abner$^{38}$ presented eight desiderata for such modifications of SQT.
CSL fared fairly well under Abner's sieve, except for desideratum d:
\bigskip
	{\tenit If a stochastic dynamical theory is used to account for the outcomes of a measurement,
it should not permit excessive indefiniteness of the outcome, where ``excessive'' is defined
by considerations of sensory discrimination.} {\tenrm This desideratum tolerates outcomes in which 
the apparatus variable does not have a sharp value, but it does not tolerate ``tails" which
are so broad that different parts of the range of the variable can be discriminated by the senses,
even if a very low probablity amplitude is assigned to the tail.}
\bigskip
	In what follows I would like to address Abner's criticism.  Essentially, I believe he has
succumbed to my disease, of viewing a new theory in an old language. As a new language appropriate
to the new theory I want to suggest a notion of stuff that resides in real space, not
configuration space.  I will distinguish three kinds of spacetime reality:
objective, projective and subjective.  

\vskip 1cm
\leftline {\title 5.1. Objective Reality}
\bigskip
	In spite of the fact that my name is on the papers$^{37, 39}$ which introduce the definition
of objective reality given here, it was really GianCarlo and Renata Grassi who saw
the need for this in relativistic CSL.  Indeed, one definition the dictionary gives
of ``objective" is ``real," so until we discuss the relativistic case
it may seem that this subsection has a redundant title.

	We begin by asking, when does a state $|\psi >$ possesses the property $A=a$?  Here I use $A$
to represent a quantity like charge (sometimes $A$ will represent
the associated operator), and $a$ stands for its value, e.g., $2e$. In SQT we
have the criterion

$$<\psi |P_{a}|\psi >\equiv <\psi |(|a><a|)|\psi >=1\eqno (5.1)$$

\noindent(here and hereafter we are taking
the statevector to be normalized to 1).  Then in fact $|\psi >= |a>$, 
and a measurement of $A$ is {\it sure} to give $a$. I think
no one would object if we
then say that $a$ is the objective value of $A$ for the state  $|\psi >$.

	This is all very well, but this definition of objective
reality is global, not local.  While, above all, it is the correspondence 
\bigskip
\hskip 1 in reality\qquad $\longleftrightarrow$\qquad statevector in Hilbert space
\bigskip
\noindent which captures the fundamental ontology behind collapse models, 
we live in spacetime, not Hilbert space. What is the reflection of that reality in spacetime?

	So let us specialize to quantities $A^{V}$ contained in a volume $V$, 
such as particle number (of any type), 
mass, charge, spin, energy, momentum, angular momentum, etc.
For example, consider the particle-number-in-$V$ operator

$$A^{V}\equiv \int_{V}d{\bf x}\xi ^{\dagger}({\bf x})
\xi ({\bf x})=\sum_{n=0}^{\infty} nP_{n}^{V}  \eqno (5.2)$$

\noindent where $P_{n}^{V}$ is the projection operator on n particles in $V$:

$$P_{n}^{V}\equiv {1\over n!}\int_{V}\cdots\int_{V}d{\bf x}_{1}\cdots d{\bf x}_{n}
\xi ^{\dagger}({\bf x}_{1})\cdots \xi ^{\dagger}({\bf x}_{n})|0>_{V}\null_{V}
\negthinspace\negthinspace<0|
\xi ({\bf x_{1}})\cdots \xi ({\bf x_{n}})\otimes1_{\bar V}\eqno (5.3)$$

\noindent ($|0>_{V}$ is the vacuum state in $V$, and $1_{\bar V}$ is the identity operator
outside of $V$). We immediately encounter a difficulty with our definition of objective
reality.  

	Let V be a sphere of radius $10^{-6}$ cm, and suppose that a hydrogen atom lies
with its nucleus at the center of the sphere. Let us ask the question: is the electron in V?
We readily calculate that $<\psi |P_{1}^{V}|\psi>\approx 1-10^{-169}$.  I would like to
say that the electron is in V (after all, even if one makes measurements every
picosecond for the age of the universe, the probability is 
miniscule that the electron will be found outside of V) but the criterion (5.1) doesn't allow it.
It is not unreasonable to say that (5.1) is unreasonable, and so we adopt a new criterion
for $a$ to be an objectively real value of $A^{V}$ for the state $|\psi >$:

$$ <\psi |P_{a}^{V}|\psi \geq 1-\epsilon \eqno (5.4)$$

This deserves some discussion.  One may feel uneasy that a criterion for spacetime
{\it objective} reality
depends upon the (exceedingly small) parameter $\epsilon$ whose value however is to be determined by 
{\it subjective} agreement. But isn't this true of everything we call real, indeed isn't it 
the best that can be done? I say the dish is in the sink (or, more generally, 
assert the location of anything), even though quantum theory and/or statistical physics 
predicts a tiny chance that when one looks it will be found elsewhere.  
I say the egg is broken (or, more generally, assert
that anything has occurred) even though there is
a tiny chance that when one looks 
not only will the egg be whole, but all records of its breakage will be blank.  

	Indeed, as Abner suggests, the only reality we are able to account
for, or should be required to account for, in a scientific theory 
is the reality apparent to the human senses and the human experience.
Then it appears proper that the criterion for spacetime reality requires human agreement,
to conform to that experience.
I regard the tail
as a tiny bit of noise in the presence of a large signal, and we have many examples
of how the human system---indeed, any device---only sees the signal in such cases. Perhaps when we have a better
understanding of the brain and its possible dependence upon quantum behavior$^{20}$ we will 
understand more deeply how a brain in a superposed state does or does not perceive reality.  
But until then, I do not think it is reasonable, and worthy of a 
desideratum, that we should expect more of the human apparatus than of any other apparatus
in terms of its ability to detect noise in the presence of a signal.

\vskip 1cm
\leftline {\title 5.2. Stuff and Projective Reality}
\bigskip

 	I believe that objective reality is not sufficient to capture
the full reality content of CSL.  Consider an electron 
in the state $|\psi>=(1/\sqrt 2)|L>+(1/\sqrt 2)|R>$, where the 
wavefunction for $|L>$ is nonzero only inside a volume $L$, and the 
wavefunction for $|R>$ is nonzero only inside a different nonoverlapping volume $R$.  Then
 
 $$<\psi |P_{0}^{L}|\psi >=<\psi |P_{0}^{R}|\psi >=<\psi |P_{1}^{L}|\psi >
 =<\psi |P_{1}^{R}|\psi >={1\over 2}\eqno (5.5)$$
 
\noindent so neither particle number 0 or 1 inside $L$ or $R$ is an objective value for $|\psi>$.
Nonetheless, I want to be able
to say that {\it something} is {\it really} inside $L$ and $R$. So I am going to define
\bigskip
$\hskip 1 in <\psi |P_{a}^{V}|\psi >$ is the projective amount of $A^{V}=a$-stuff in $V$ 
\bigskip
\noindent and regard stuff as (projectively) real.  This is another kind of reality
than objective reality.  The latter is what is produced in the usual von Neumann
strong coupling impulsive measurements, designed so that, if collapse did not take place, 
the apparatus is forced 
into a superposition of macrostates, each corresponding to a different
eigenstate of operator $A^{V}$, one of which
collapse makes objectively real.  While I don't believe that a theoretical 
notion of reality {\it has} to include a method of experimental verification, it doesn't hurt that 
Aharonov and Vaidman$^{40}$ have shown that stuff 
can also be measured in certain circumstances by so-called protective weak coupling measurements,
where the apparatus is never forced into a superposition of macrostates: it just indicates the
projective value.  

	In what follows I will specialize to particlenumber $n$-stuff in $V$ which for short,
and if there is no chance of confusion, I will
sometimes call particlenumber stuff, or $n$-stuff in $V$, or $n$-stuff, or even just stuff:
	
$$<\psi |P_{n}^{V}|\psi >={N\choose n} 
\int_{V}d{\bf x}_{1}\cdots\int_{V}d{\bf x}_{n}\int_{\bar V}d{\bf x}_{n+1}
\cdots\int_{\bar V}d{\bf x}_{N}|\psi({\bf x}_{1},\dots {\bf x}_{N}|^{2}\eqno (5.6)$$
	
\noindent (it is assumed that there are $N$ particles of this type in the world, and
I have used the symmetry of the wavefunction to simplify).  One could argue
that this is the most important kind of stuff for human beings because
our pictures of what the world looks like are literally shaped 
by the particle densities that surround us. 
But stuff for other variables can be discussed.  Indeed, 
there is nothing to prevent one from discussing stuff associated with noncommuting
variables (stuff is just a c-number), e.g., particlenumber stuff 
and momentum stuff are simultaneously in $V$.  
Indeed, it is a big relief to think of these quantities as
actually being there together. This replaces the notion of ``propensities" of Heisenberg which
always caused me consternation: I find it hard to think of something that
{\it might} show up in the future
{\it if} the right experiment just happens to be performed, as having some real
status.

	When the $n$-stuff in $V$ achieves the value 1-$\epsilon$, then the projective 
reality becomes objective.  Thus, instead of the statement ``so much particlenumber $n$-stuff
is in $V$" one may then say that ``$n$ particles are in $V$."  It is important and interesting
to consider the differences between stuff and particle number.  For example, the sum of
0-stuff+1-stuff+\dots +N-stuff in any $V$ is conserved (=1), but the sum of particles in $V$
is not conserved.  Conversely, if one divides all of space into cells, the sum of e.g., 1-stuff
in all cells is not conserved. It can range from a maximum of $N$ (if, for
example, the wavefunction is such that each particle is in a different cell) to zero (if, for 
example, each cell happens to hold two or more particles). However, the particle number in 
all cells is conserved (assuming fermions with no particle creation). 

	I think that the most interesting aspect of the difference lies in the flow in spacetime.  
For example, consider $N$=2. It is easily found from the evolution equation (4.1) that

$$\eqalign{&{\partial \over \partial t}|\psi({\bf x}_{1},{\bf x}_{2},t)|^{2}
+{\bf \nabla}_{1}\cdot{\bf J}_{1}+{\bf \nabla}_{2}\cdot{\bf J}_{2}=\cr
&\qquad |\psi({\bf x}_{1},{\bf x}_{2},t)|^{2}\biggl[ F({\bf x}_{1},{\bf x}_{2},t)-
\int\int_{V+\bar V}d{\bf z}_{1}d{\bf z}_{2}F({\bf z}_{1},{\bf z}_{2},t)
|\psi({\bf z}_{1},{\bf z}_{2},t)|^{2}\biggr]}\eqno (5.7)$$

\noindent ($|\psi|^2$ here is normalized to 1).  It doesn't matter for our argument, but for completeness
sake we exhibit F:

$$ F({\bf x}_{1},{\bf x}_{2},t)=2\int d{\bf z}w({\bf z},t)
[G({\bf z}-{\bf x}_{1})+G({\bf z}-{\bf x}_{2})-4\lambda \Phi({\bf x}_{1}-{\bf x}_{2})]\eqno(5.8a)$$

$$G({\bf z}-{\bf x})={1\over (2\pi a^{2})^{3/4}}e^{-{1\over 2a^{2}}({\bf z}-{\bf x})^{2}},\qquad
\Phi({\bf x}-{\bf y})=e^{-{1\over 4a^{2}}({\bf x}-{\bf y})^{2}}\eqno(5.8b,c)$$

By appropriately integrating Eq.(5.7) over $V$ and $\bar V$, we obtain

$$\eqalign{&{\partial \over \partial t}<\psi ,t |P_{1}^{V}|\psi ,t >+
\int_{\Sigma_{V}}d{\bf \Sigma}_{1}\cdot\biggl[\int_{\bar V}-\int_{V}\biggr]d{\bf x}_{2}{\bf J}_{1}+
\int_{\Sigma_{V}}d{\bf \Sigma}_{2}\cdot\biggl[\int_{\bar V}-\int_{V}\biggr]d{\bf x}_{1}{\bf J}_{2}
=\cr
&\biggl[\int_{V}\int_{\bar V}+\int_{\bar V}\int_{V}-
<\psi ,t |P_{1}^{V}|\psi ,t >\int\int_{V+\bar V}\biggr]d{\bf x}_{1}d{\bf x}_{2}
F({\bf x}_{1},{\bf x}_{2},t)|\psi({\bf x}_{1},{\bf x}_{2},t)|^{2}}\eqno(5.9)$$ 

\noindent Now, if $<\psi ,t |P_{1}^{V}|\psi ,t >=1$, the source term (the second line in Eq.(5.9))
vanishes: there is then one particle in $V$ and one outside of $V$, so  
$|\psi({\bf x}_{1},{\bf x}_{2},t)|^{2}$ vanishes unless
one argument lies in $V$  and the other lies in $\bar V$, and the integrals cancel.  
(If $<\psi ,t |P_{1}^{V}|\psi ,t >=1-\epsilon$, the source term is still
negligibly small). Moreover, the surface integral in (5.9) will vanish, at least
for a while, if the stuff of either particle is far from the boundary.
	Thus when reality is {\it objective},
Eq.(5.9) describes the particle number's  conserved local
 flow in spacetime, at least for a while (which may be 
 prolonged by moving or distorting V).  On the other hand, when reality is 
{\it projective}, the 1-stuff flows nonlocally as well as locally.
  
	The reason for this behavior is that collapse narrows wavefunctions in spacetime.
For example, if the whole wavefunction of a single particle
is contained in $V$, the collapse does not transfer any of it out of $V$. But if the wavefunction
lies partially in and partially out of $V$, there is a transfer regardless of the extension
of the wavefunction.  It is this nonlocal flow of stuff that makes a collapse theory violate Bell's 
inequality and agree with experiment and SQT.

	This allows some insight into the difference between classical behavior
 and quantum behavior.
 
  The local flow of an objectively real particle number
 from one volume of space to an objectively real particle number in an adjacent volume is
 what classical physics is about. An objective number of particles
 (of every type) in a macroscopic object can be imagined wrapped in
 a volume $V$ which moves like a rigid object in spacetime.  Even 
 a microscopic particle, as long as it moves freely, can be surrounded by a large enough
 or growing volume so that its particle number, charge, energy etc. are objective and
 flow locally in spacetime.
 
 	However, particle number can flow into stuff, and
that is what quantum theory is about. For example, when a 
measurement with different possible outcomes
takes place, 
many particles of the macroscopic apparatus move into different volumes of space. 
For another example, if a single particle scatters
or if it spreads outside of $V$, no longer is $<\psi ,t |P_{1}^{V}|\psi ,t >=1-\epsilon$.
 In these
cases, particle number is replaced by stuff
and projective reality has replaced objective reality. 

Of course, quantum theory
is also about stuff flowing into stuff.
 
	  It is only the collapse dynamics which allows stuff to flow into 
particle number, for example, when the collapse
after a measurement takes place. This is the only mechanism which
permits projective reality to become objective.  Without collapse, I venture to say,  
objective
reality would eventually disappear, and all reality would 
eventually become projective.

	This discussion may be summarized by the following diagram:
\bigskip
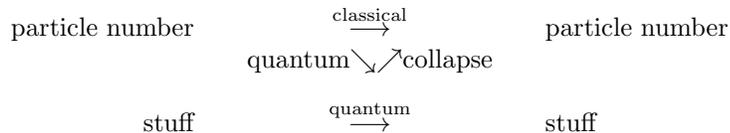
\moveright 1 in\vbox{\halign{\hfil#&\qquad\hfil#\hfil&\qquad#\hfil\cr
particle number&$\buildrel{\rm classical}\over\longrightarrow$&particle number\cr
\null&{\tenrm quantum}$\searrow$$\nearrow${\tenrm collapse}&\null\cr
\null&\null&\null\cr
 stuff&$\buildrel{\rm quantum}\over\longrightarrow$&stuff\cr}}
\vskip 1cm
\leftline {\title 5.3. Relativity and Objective Reality}
\bigskip
	In this subsection we will see
that the ``peaceful coexistence" (one of many apt and useful terms coined
by Abner) 
of relativity and a collapse model requires a 
new category of reality, and a redefinition of an old one.
A relativistic version of
CSL has been constructed, with some successes and some problems.$^{36, 37, 39, 41, 42}$
The main point relevant here is that 
collapse in {\it each} Lorentz reference frame proceeds essentially 
just like nonrelativistic collapse.

	The consequence of this for us can be made clear by the following example.  Consider
a particle in reference frame $o$ which, for $t<T$, is in a superposition
of two equal amplitude spatially separated packets with worldlines at $L$ and $R$.
Therefore, for $t<T$, the 1-stuff on each
worldline is equal to 1/2. Call $\alpha$ some such small spacetime neighborhood on $R$.  Suppose
that at time $t=T$ a particle-detecting apparatus situated near $R$ is turned on, and
the result is that the particle is not detected. Therefore, in this frame, for
$t>T$, the 1-stuff is $1-\epsilon$ along $L$
and the 0-stuff is 1-$\epsilon$ along $R$, so particlenumber has the objective
values 1 along $L$, and 0 along $R$. Call $\beta$ some such small spacetime neighborhood on $L$.
	
		Now consider a different frame $o'$ which, at time $t'=T'$, passes through
$\beta$ and $\alpha$.  In this frame, the measurement has not yet begun, so the 1-stuff
at $\beta$ and $\alpha$ is equal to 1/2.  But this means that,
the two frames differ concerning the amount of stuff at $\beta$: according to our definitions,
frame $o$ says that that the reality is objective at $\beta$ while frame $o'$ says it is projective.

	So the first lesson we learn is that the amount of stuff in
a spacetime region can be frame-dependent. Of course, 
this is not a new thing in Relativity: the length of a stick, the 
time interval on a clock, the energy of a particle, etc., all are frame-dependent.
Considering that, in collapse models, reality corresponds to 
a statevector in Hilbert space, and
that the statevector is frame-dependent, it should not be surprising (although
it may be disappointing to preconceived hopes)
that reality can be frame-dependent.  Yet, although reality {\it can} be frame-dependent, 
it doesn't {\it have} to be frame-dependent.  I think that the circumstances in which
it is not frame-dependent, first really appreciated by GianCarlo, is quite satisfying.

	We are led to define three kinds of reality in a relativistically
invariant way, contingent upon what is seen in {\it all} reference frames
which pass through a spacetime region.  If in all frames the stuff is projective, 
then we will say that reality is projective in that region.  
If in all frames the criterion (5.4) is satisfied,  
then we will say that reality is objective in that region.  If in some frames the
stuff does not satisfy (5.4) and in some other frames it does, we will say that reality
is subjective (= frame-dependent) in that region.

	Now we note that, in the example above, objective reality exists  
only on and within the forward light cone of the measurement region (on the 
$R$ worldline immediately after the measurement, but on the $L$
worldline only beyond where it cuts this light cone).  That is, 
the message 
from a measuring instrument which calls objective reality into existence
 travels
no faster than the speed of light. Isn't that nice?

	A few more comments.
	  
	The collapse location (where a packet grows or diminishes) 
can be objective (agreed upon by all observers) or subjective (not agreed upon).
In the above example, the location on $R$
is objective and the location on $L$ is subjective.  The 
collapse location is not a thing that can be measured, so there
 is no need for this location to
be objective.

	In the above example, all frames agree that
the cause of the collapse and the subsequent emergence of
objective particlenumber at $L$ was due to
the apparatus being turned on at $R$.  But the cause of
collapse is not always necessarily objective
as it is in this case.  Consider a situation
in which there are two apparatuses, one at $L$ as well as one at $R$, 
and where both are turned on simultaneously in one frame. This leads one to conclude
that the cause of collapse in some frames is the turning on of the apparatus at $L$, 
and in other frames it is the turning on of the apparatus at $R$. The cause of
collapse is not a thing which can be measured, so there
is no need for it to be objective.

	Finally, I want to give one more reason for tails: I can't
see how to make a relativistic theory without them. If you have a tail, no matter how small, 
and you know the field $w({\bf x},t)$ which the state vector evolved under, you can run
the evolution equation (4.1) backwards and recover the statevector at any earlier time. 
If on the other hand, the tail was completely cut off, you get a nonsensical irrelevant
earlier statevector, even in SQT.  ``So what?" you say. ``There 
is no need to run a statevector backwards since time runs forward and one can't stop that."
Well, but one {\it can} go to another reference frame, and in so doing
the frame sweeps backwards in time.  I cannot see how you could get sensible results
in another Lorentz frame without having the tail to tell you how to do it.  And, this
ends the tale I wish to tell.
\vskip 1cm
\centerline{\big{Acknowledgments}}
\bigskip   Even though in this paper
I have disagreed with Abner, 
I will always
listen to him (see (0.1)) with great pleasure, as I have for three decades.   
I would like to thank Abner for being what he is.

	I would also like to thank GianCarlo Ghirardi and Renata Grassi, who have
recently completed their own insightful endorsement of tails,$^{43}$ for
animated and informative conversations earlier this year (my fourth Sabbatical). 
This, along with Abner's comments on tails, stimulated this paper. 
In addition, I wish to thank the Hughes
Foundation for a grant supporting this work.

\bigskip	
\centerline {\big{References and Remarks}}
\bigskip
\noindent 1. P. Pearle, Amer. Journ of Phys. {\bf 35}, 742 (1967).

\noindent 2. Abner has since reminded me of 
a role I played in this work.  I referred him to a good friend of mine 
still at Harvard, Joe Snider, who was to be found at the bottom of 
the elevator shaft in the Jefferson laboratory working
on the Pound-Rebka-Snider experiment, shooting photons 
up the shaft to measure their gravitational red shift.  Joe directed Abner
to some photon counting
experiments which had already been performed and which proved helpful.

\noindent 3.  J. F. Clauser, R. A. Holt, M. A. Horne and A. Shimony,
 Phys. Rev. Lett {\bf 23}, 880 (1969).

\noindent 4. P. Pearle, Phys. Rev. {\bf D10}, 526 (1974).

\noindent 5. J. F. Clauser and M. Horne, Phys. Rev. {\bf D2}, 1418 (1970).

\noindent  6. N. D. Merminin {\it New Techniques 
in Quantum Measurement Theory}, edited by D. M. Greenberger 
(N.Y. Acad. of Sci., N.Y., 1986), p. 422.

\noindent 7. D. Bohm and J. Bub, Reviews of Modern Physics {\bf 38}, 453 (1966).

\noindent 8. P. Pearle, Foundations of Physics {\bf 12}, 249 (1982).

\noindent 9. P. Pearle, Journal of Statistical Physics {\bf 41}, 719 (1985)

\noindent 10. P. Pearle, Physical Review D{\bf 33}, 2240 (1986)

 \noindent 11. P. Pearle, in {\it The Wave-Particle Dualism},
 edited by S. Diner et. al (Reidel, Dordrecht 1984), p. 457.

\noindent 12. I. Prigogine, {\it Non-equilibrium Statistical Mechanics}
 (Interscience, New York, 1962).
 
\noindent 13. S.Chandrasekhar, Revs. Mod. Phys. {\bf 15}, 1 (1943).

\noindent 14. P. Pearle, Physical Review {\bf D13}, 857 (1976).

\noindent 15. E. Wong, {\it stochastic Processes in Information and Dynamical Systems},
(McGraw-Hill, New York 1971).

\noindent 16. P. Pearle, International Journal of Theoretical Physics {\bf 48}, 489 (1979)

\noindent 17. P. Pearle, Physical Review {\bf A48}, 913 (1993).

\noindent 18.  P. Pearle, preprint "Wavefunction collapse models with nonwhite noise," (1994).

\noindent 19. F. Karolyhazy, Nuovo Cimento {\bf 42}A, 1506 (1966);
 F. Karolyhazy, A Frenkel and B. Lukacs in {\it Physics as Natural Philosophy}, 
edited by A. Shimony and H. Feshbach (M.I.T. Press, Cambridge 1982), p. 204;
in {\it Quantum Concepts in Space and Time}, edited by R. Penrose 
and C. J. Isham (Clarendon, Oxford 1986), p. 109; A. Frenkel, Found. Phys. {\bf 20}, 159 (1990).

\noindent 20. R. Penrose in {\it Quantum Concepts in Space and Time}, 
edited by R. Penrose and C. J. Isham (Clarendon, Oxford 1986), p. 129; {\it The Emperor's
New Mind}, (Oxford University Press, Oxford, 1992).

\noindent 21. L. Diosi, Phys. Rev. A{\bf 40}, 1165 (1989).

\noindent 22. G. C. Ghirardi, R. Grassi and A. Rimini, Phys. Rev. A{\bf 42}, 1057 (1990).

\noindent 23. P. Pearle and E. Squires, Phys. Rev. Lett. {\bf 73}, 1 (1994).

\noindent 24. P. Pearle, in {\it New Techniques 
in Quantum Measurement Theory}, edited by D. M. Greenberger 
(N.Y. Acad. of Sci., N.Y., 1986), p. 539. 

\noindent 25. P. Pearle, in {\it Fundamental Questions in Quantum Mechanics},
 edited by L. M. Roth and A. Inomata (Gordon and Breach, New York, 1984), p.41.

\noindent 26. P. Pearle in {\it Quantum Concepts in Space and Time}, 
edited by R. Penrose and C. J. Isham (Clarendon, Oxford 1986), p. 84.

\noindent 27. G. C. Ghirardi, A. Rimini and T. Weber, Physical Review D {\bf 34}, 470 (1986);
Physical Review D{\bf 36}, 3287 (1987); Foundations of Physics {\bf 18}, 1, (1988)

\noindent 28.  J. S. Bell in {\it Schrodinger-Centenary celebration of a polymath}, 
edited by C. W. Kilmister (Cambridge University Press, Cambridge 1987) 
and in Speakable and unspeakable in quantum mechanics, 
(Cambridge University Press, Cambridge 1987), p. 167.

\noindent 29. P. Pearle, Physical Review A {\bf 39}, 2277 (1989).

\noindent 30. G. C. Ghirardi, P. Pearle and A. Rimini, Physical Review A{\bf 42}, 78 (1990).

\noindent 31. N. Gisin, Helvetica Physica Acta {\bf 62}, 363 (1989).

\noindent 32. V. P. Belavkin, Physics Letters A{\bf 140}, 355 (1989).
 
\noindent 33. L. Diosi, Physics Letters A{\bf 132}, 233 (1988).

\noindent 34. J. S. Bell in {\it Sixty-Two Years of Uncertainty},
edited by A. Miller (Plenum, New York 1990), p. 17.

\noindent 35. G. C. Ghirardi and A. Rimini in {\it Sixty-Two Years of Uncertainty}, 
edited by A. Miller (Plenum, New  York 1990), p. 167.
 
\noindent 36. P. Pearle in {\it Sixty-Two Years of Uncertainty},
edited by A. Miller (Plenum, New York 1990), p. 193.

\noindent 37. G. C. Ghirardi and P. Pearle in {\it Proceedings of the Philosophy
 of Science Foundation 1990, Volume 2}, edited 
 by A. Fine, M. Forbes and L. Wessels (PSA Association, Michigan 1992), p. 19 and p. 35. 

\noindent 38. A. Shimony in {\it Proceedings of the Philosophy
 of Science Foundation 1990, Volume 2}, edited 
 by A. Fine, M. Forbes and L. Wessels (PSA Association, Michigan 1992), p. 49. 

\noindent 39. G. C. Ghirardi, R. Grassi and P. Pearle, Foundations of Physics {\bf 20}, 1271 (1990):
in {\it Proceedings of the Symposium on the Foundations of Modern Physics 1990},
edited by P. Lahti and P. Mittelstaedt, (World Scientific, Singapore 1991), p. 109.

\noindent 40. Y. Aharonov and  L. Vaidman, Phys. Rev. {\bf A41}, 11 (1990), aand in this volume.

\noindent 41. P. Pearle, in {\it Quantum 
Chaos-Quantum Measurement}, edited by P. Cvitanovic et. al., (Kluwer, the Netherlands 1992), p. 283.

\noindent 42  G.C. Ghirardi, R. Grassi, J. Butterfield and G. N. Fleming, Foundations
of Physics {\bf 23}, 341 (1993).

\noindent 43. G. C. Ghirardi, R. Grassi and F. Benatti, preprint IC/94/107,
"Describing the Macroscopic World: 
Closing the Circle Within the Dynamical Reduction Program," (1994).

\bye